\documentclass[12pt,preprint]{aastex}

\usepackage{lscape}

\newcommand{\letter}{{\it Letter}}
\newcommand{\paperone}{Paper~I}
\newcommand{\figname}{Figure~}
\newcommand{\tabname}{Table~}
\newcommand{\eqnname}{Eqn.~}

\newcommand{\sdssu}{$u$}
\newcommand{\sdssg}{$g$}
\newcommand{\sdssr}{$r$}
\newcommand{\sdssi}{$i$}
\newcommand{\sdssz}{$z$}
\newcommand{\jhni}{$I$}
\newcommand{\effwavesdssz}{8931~\AA}
\newcommand{\effwavejhni}{8062~\AA}
\newcommand{\redshiftupperlimit}{1.6}

\newcommand{\reffm}{r_{\rm eff}}
\newcommand{\rfiber}{$r_{\rm fiber}$}
\newcommand{\rcircle}{$r_{\rm circle}$}
\newcommand{\rcirclem}{r_{\rm circle}}

\newcommand{\hiiregion}{\ion{H}{2}~region}
\newcommand{\hiiregions}{\ion{H}{2}~regions}
\newcommand{\lineids}{H$\alpha$, H$\beta$, H$\gamma$,
  $\left[\right.$\ion{O}{2}$\left.\right]$, $\left[\right.$\ion{N}{2}$\left.\right]$}
\newcommand{\halpha}{H$\alpha$}
\newcommand{\hbeta}{H$\beta$}
\newcommand{\hgamma}{H$\gamma$}
\newcommand{\oii}{$\left[\right.$\ion{O}{2}$\left.\right]$}
\newcommand{\nii}{$\left[\right.$\ion{N}{2}$\left.\right]$}
\newcommand{\halpham}{{\rm H}\alpha}

\newcommand{\oiim}{\left[{\rm O\,II}\right]}

\newcommand{\lambdaoii}{$\lambda3728$}

\newcommand{\sfr}{SFR}
\newcommand{\sfrmath}{\mathrm{SFR}}

\newcommand{\ba}{b/a}
\newcommand{\ergs}{erg~s$^{-1}$}

\newcommand{\redshiftrange}{0.065 to 0.075}
\newcommand{\magnituderange}{-19.5 to -22}
\newcommand{\baedgeon}{0.17}

\newcommand{\rfibersdss}{1.5\arcsec}
\newcommand{\solarmassperyr}{M_{\odot}~\mathrm{yr}^{-1}}
\newcommand{\spectralresolution}{70~km~s$^{-1}$}

\newcommand{\sfrscaleha}{1.27 \times 10^{41}~\mathrm{erg~s}^{-1}}
\newcommand{\sfrscaleoii}{7.14 \times 10^{40}~\mathrm{erg~s}^{-1}}
\newcommand{\fluxscale}{$10^{40}~\mathrm{erg~s}^{-1}$}
\newcommand{\scalelengthratiohalphaIbandscatter}{0.52}
\newcommand{\scalelengthratiohalphaIbandmean}{1.33}
\newcommand{\scalelengthratiohalphaIbandsd}{0.08}

\newcommand{\changeinscalelengthratiohalphaIband}{$\approx$30\%}
\newcommand{\fluxfiber}{Fiber}
\newcommand{\fluxcircle}{Circle}
\newcommand{\fluxtotal}{Total}
\newcommand{\luminositydropwithinclination}{three}
\newcommand{\relextHalpha}{1.2~mag}
\newcommand{\reffsampledbyfiber}{0.5}
\newcommand{\galsimaxisratio}{0.15}
\newcommand{\galsimfibertocircle}{2.7}
\newcommand{\zbandfaceonreff}{2.89\arcsec}
\newcommand{\fibertototalHaLum}{14.0\%}
\newcommand{\fibertototalHaLumfrac}{0.14}
\newcommand{\oiitohalpha}{about 0.4}
\newcommand{\oiitohalphaUstoHopkins}{about a factor of two}
\newcommand{\oiitohalphaHopkins}{in which the value is 0.23}
\newcommand{\oiitohalphaKennicutt}{in which the value is 0.4}
\newcommand{\sfraverage}{about 1.1~$\solarmassperyr$}
\newcommand{\changeinfibertocircle}{less than $5\%$}

\begin{document}

\title{Extinction in Nebular Luminosities \& Star Formation
  Rate of Disk Galaxies: Inclination Correction}

\author{Ching-Wa Yip\altaffilmark{} \& Alex~S.~Szalay\altaffilmark{}}

\altaffiltext{}{Department  of  Physics   and  Astronomy,  The  Johns
  Hopkins      University,     Baltimore,     MD      21218,     USA}

\email{cwyip@pha.jhu.edu; szalay@jhu.edu}

\begin{abstract}
Star  formation is  one  of  the most  important  processes in  galaxy
formation.    The   luminosity   of   \halpha  \   recombination   and
\oii\lambdaoii  \  forbidden  emissions  remain  to be  most  used  in
measuring formation rate of massive stars in galaxies.  Here we report
the    inclination     dependency    of    continuum-subtracted    and
aperture-corrected   nebular  luminosities,  including   \lineids,  of
disk-dominated  galaxies in  the local  universe.   Their luminosities
decrease by a factor  of \luminositydropwithinclination \ from face-on
to  edge-on  (axis  ratio   limit  =  \baedgeon)  orientations.   This
dependence is deduced  to be caused by extinction  due to diffuse dust
within   the  disks   with   an  amplitude   of  \relextHalpha.    The
line-luminosity--inclination relation  provides a novel  way to remove
extinction in emission  lines and present star formation  rate of disk
galaxies out to redshift of \redshiftupperlimit.
\end{abstract}


\section{Motivation} \label{section:motivation}

Star  formation is  one of  the  most important  phenomena during  the
formation of a galaxy.  To measure the star formation rate (\sfr), the
luminosities of both \halpha \ and \oii \ are used because they depend
on   the   ionizing    photon   luminosity   \citep[e.g.,   \eqnname~2
of][]{1998ARA&A..36..189K}. The radiation from massive stars, taken to
be  a blackbody  \citep{1927ApJ....65...50Z} and  determined  to arise
from O/B  stars, is the  ionizing source to  the \hiiregion \  gas, or
nebular gas.  The  \sfr \ have been quantified  using large samples of
galaxies   \citep[e.g.,][]{2004MNRAS.351.1151B}   and  across   cosmic
distances
\citep[e.g.,][]{1996ApJ...460L...1L,1997ApJ...486L..11C,1998ApJ...498..106M,2006ApJ...651..142H}.
On    average,     star-forming    galaxies    are     dusty    (e.g.,
\citet{2000ApJ...533..682C,2007MNRAS.379.1022D,2007ApJ...659.1159S,2007AJ....134.2385H,2008ApJ...687..976U,2008ApJ...681..225B,2009ApJ...691..394M,2010MNRAS.404..792M,2010ApJ...718..184C,2011MNRAS.tmp.1545W}
and   \citet[][hereafter   \paperone]{2010ApJ...709..780Y}).   It   is
therefore critical to get \sfr \ free from extinction.

The extinction correction to  the spectral energy distribution of disk
galaxies can  be decomposed into two  terms: the inclination-dependent
and        -independent       (e.g.,        face-on)       corrections
\citep[see][]{2008ApJ...678L.101D}.  While  the inclination dependency
of  optical   stellar  continuum  in  disk   galaxies  is  established
\citep{2007MNRAS.379.1022D,2008ApJ...687..976U,2008ApJ...681..225B,2009ApJ...691..394M,2010ApJ...709..780Y},
that of  individual nebular emissions  is still lacking.  The  goal of
this  \letter  \  is   to  determine  the  inclination  correction  to
individual nebular  emissions and therefore  \sfr \ of  disk galaxies.
We  quantify  the  luminosity   of  the  \halpha,  \hbeta,  \hgamma  \
recombination  lines, and  the  \oii,  \nii \  forbidden  lines, as  a
function of inclination for a sample of local disk-dominated galaxies.
The  line-luminosity--inclination  relation  can  be  used  to  remove
inclination-dependent extinction of disk galaxies.

\section{Analysis} \label{section:data}

Our sample is  composed of galaxies from the  Sloan Digital Sky Survey
\citep[SDSS,][]{2000AJ....120.1579Y}        Data       Release       6
\citep[DR6,][]{2008ApJS..175..297A}  that   are  morphologically  disk
dominated and spectroscopically  star forming.  The selection criteria
were  described   in  detail  in   Paper~I.   To  summarize   the  key
characteristics,   the   sample   is   (1)   within   redshifts   from
\redshiftrange; (2)  within \sdssr-band Petrosian  absolute magnitudes
from \magnituderange;  (3) volume  limited, in avoiding  the Malmquist
bias  and  (4)  showing  a  uniform distribution  of  inclination,  in
avoiding bias in  galaxy properties from one inclination  to the next.
Following \paperone,  the inclination  of a galaxy  is proxied  by its
\sdssr-band  axis  ratio,  with  the consideration  that  the  average
ellipticity is small \citep[16\%,][]{2004ApJ...601..214R}.

To quantify the line luminosity as a function of inclination, we first
construct  high  signal-to-noise  composite  spectrum, at  a  spectral
resolution   of   \spectralresolution,    in   bins   of   inclination
(\tabname\ref{tab:line}).   We then  measure  the continuum-subtracted
line luminosities  in the composites.  The  methodologies in continuum
estimation and  line fitting were discussed in  \paperone. Finally, we
apply  an aperture  correction to  obtain line  luminosities  that are
expected  from the  whole  galaxies.  The  correction is  specifically
developed in  this work  for the purpose  of inclination  studies with
fiber spectroscopy. 

The SDSS spectra  of the galaxies are downloaded  through the Spectrum
Services \citep{2004ASPC..314..185D}.  Line luminosities are expressed
in \fluxscale, and the vacuum wavelength convention is used throughout
the  analysis.   The flux-to-luminosity  conversion  made  use of  the
cosmological parameters $\Omega_{V} = 0.73, \Omega_{M} = 0.27$, and $h
= 0.71$.

\subsection{Aperture Correction}

The SDSS spectroscopic  fiber is \rfibersdss \ in  radius (\rfiber \ =
 \rfibersdss), which measures  \reffsampledbyfiber \ half-light radius
 area   on   our  disk   galaxies   (see  \tabname\ref{tab:size}   and
 \S\ref{section:fiber-to-circle}).  A  two-step procedure is developed
 to obtain line  luminosities of the whole galaxies  from the observed
 luminosities  through  the fiber.   First,  the  fiber luminosity  is
 converted to the central luminosity within a radius, \rcircle, on the
 plane     of     a    galaxy     (hereafter,     the    step     {\it
 \fluxfiber-to-\fluxcircle}).   Second,  this  central  luminosity  is
 converted  to the  luminosity  of  the whole  galaxy  (the step  {\it
 \fluxcircle-to-\fluxtotal}).

\subsubsection{Step 1: \fluxfiber-to-\fluxcircle \ Correction}\label{section:fiber-to-circle}

When  a  disk  galaxy  is  inclined, a  circular  spectroscopic  fiber
measures an  elliptical area  {\it on the  plane of the  disk}, rather
than a circular area.   The luminosity through the spectroscopic fiber
is contributed  not only from all  of the isophotes  within the radius
\rcircle  \ =  \rfiber  \ on  the  plane of  the  galaxy.  Light  from
isophotes that  lie within an ellipse  -- which is  the fiber aperture
projected on the plane of the  disk -- also contribute.  The minor and
major  axes of  this ellipse  are respectively  \rfiber \  and \rfiber
$/(\ba)$. Therefore, to obtain the central luminosity within \rcircle,
the luminosity outside  of this circle, but within  the ellipse, would
needed to be subtracted  from the fiber luminosity.  This ``luminosity
excess'' increases with inclination, and is zero for face-on galaxies.

To quantify this excess, the  radial distribution of in~situ \halpha \
luminosity    is     needed.     We    adopt     the    findings    by
\citet{1969ApJ...155..417H},      \citet{1983ApJ...267..563H}      and
\citet{1993A&AS..102..229A}.   These authors  found in  the  \halpha \
luminosity an  exponential dependence  with radius, separately  in two
disk  galaxy samples.   For the  half-light  radius of  the \halpha  \
profile, we adopt the ratio between the half-light radius of \halpha \
and     that     of     the     \jhni-band    in     field     spirals
\citep{2006AJ....131..716K}.
The ratio involving  the \jhni \ band is used because  we can then use
the half-light radius  in the SDSS \sdssz \ band  as a surrogate given
their similar  effective wavelengths, respectively  \effwavejhni \ and
\effwavesdssz. The ratio is

\begin{equation}
\frac{{r_{{\rm  eff},  {\rm  H}\alpha}}(1)}{{r_{{\rm eff},  z}}(1)}  =
\scalelengthratiohalphaIbandmean  \pm \scalelengthratiohalphaIbandsd \
,
\label{eqn:reffHaToz}
\end{equation}

\noindent
and with a  sample scatter of \scalelengthratiohalphaIbandscatter. The
 bracket  denotes  the inclination,  that  is  face-on  in this  case.
 Similar to  \paperone, we fit  to apparent size vs.   inclination the
 formula:  ${\log}_{10}  \left[  {\reffm}(\ba) \right]  =  {\log}_{10}
 \left[ {\reffm}(1) \right] - \beta \, {\log}_{10} \left( \ba \right)
\label{eqn:reffVsInclination}$. The results in \tabname\ref{tab:size}
show  that  the  apparent   size  increases  with  inclination,  found
previously         in         other         studies         \citep[][,
\paperone]{1992MNRAS.254..677H,2006A&A...456..941M,2009ApJ...691..394M},
and shown  here for the 5  SDSS bands \sdssu,  \sdssg, \sdssr, \sdssi,
\sdssz.   The  face-on half-light  radius  of  the galaxies  increases
toward shorter wavelengths, consistent  with the galaxies being dusty.
As  such, an  extinction radial  gradient  causes an  increase in  the
apparent  half-light radius;  and the  extinction amplitude  is larger
toward  shorter  wavelengths. The  average  \sdssz  \ band  half-light
radius of our galaxies, as required in \eqnname\ref{eqn:reffHaToz}, is
\zbandfaceonreff \ (\tabname\ref{tab:size}).

A numerical integration is  then used to calculate the Fiber-to-Circle
luminosity,  $R_{\mathrm{\fluxfiber-to-\fluxcircle}}(\ba)$,  where the
surface brightness  of \halpha \  follows an exponential  profile (see
\eqnname\ref{eqn:expprof}  below) with  the half-light  radius derived
above. The luminosity ratio is fully determined from the axis ratio of
the  galaxy, the \halpha  \ half-light  radius, and  the spectroscopic
fiber           radius           \rfiber.            We           find
$R_{\mathrm{\fluxfiber-to-\fluxcircle}}(\ba = \galsimaxisratio)$ to be
\galsimfibertocircle,  decreases gradually  as  inclination decreases,
and   is   unity  at   the   face-on   orientation.    The  value   of
$R_{\mathrm{\fluxfiber-to-\fluxcircle}}(\ba)$ is tested to depend only
mildly   on   the   \halpha-to-continuum  half-light   radius   ratio,
\changeinfibertocircle             \             change            for
\changeinscalelengthratiohalphaIband   \    change   in   this   ratio
(correspond to \scalelengthratiohalphaIbandmean \ $\pm$ sample scatter
from \citet{2006AJ....131..716K}).

\subsubsection{Step 2: \fluxcircle-to-\fluxtotal \ Correction}

 We calculate  in this step the \fluxcircle-to-\fluxtotal  \ \halpha \
      luminosity by  considering the total luminosity  inside a radius
      $r$ in a galaxy with exponential \halpha \ profile:

\begin{eqnarray}
L(r) & = & \int_{0}^{r} I(r) \, 2 \pi r \, dr \ , \label{eqn:expprof}
\\ & = & 2 \pi I_{0}
     {\left[ {  { \frac{e^{a  \, r}}{a} }  \, {\left( r  - \frac{1}{a}
     \right)} + { \frac{1}{a^2}} } \right]} \ ,
\label{eqn:totalflux}
\end{eqnarray}

\noindent
where $I(r) =  I_{0} \, \exp(a\, r)$, and $a  \equiv -1.68 / {{r_{{\rm
eff}, {\rm H}\alpha}}^{1}}$.  The total luminosity, $L(\infty)$, is $2
\pi I_{0} / a^{2}$.   The \fluxcircle-to-\fluxtotal \ luminosity, $L(r
=  \rcirclem   =  \rfibersdss)/L(\infty)$,  can   therefore  be  fully
determined from the  \halpha \ half-light radius and  the fiber radius
only.   The  \fluxcircle-to-\fluxtotal   \  \halpha  \  luminosity  is
determined         to         be        \fibertototalHaLum,         or
$R_{\mathrm{\fluxcircle-to-\fluxtotal}} = \fibertototalHaLumfrac$.

Combining both corrections, the aperture-corrected line luminosity is finally

\begin{equation}
L_{\mathrm{Total}}(\ba)    =    \frac{   L_{\mathrm{Fiber}}(\ba)    }{
  R_{\mathrm{Fiber-to-Circle}}(\ba)                              \times
  R_{\mathrm{Circle-to-Total}} } \ .
\end{equation}

\section{Nebular Luminosity  \& Present  Star Formation Rate as a
  function of Inclination}

\begin{figure}
\plotone{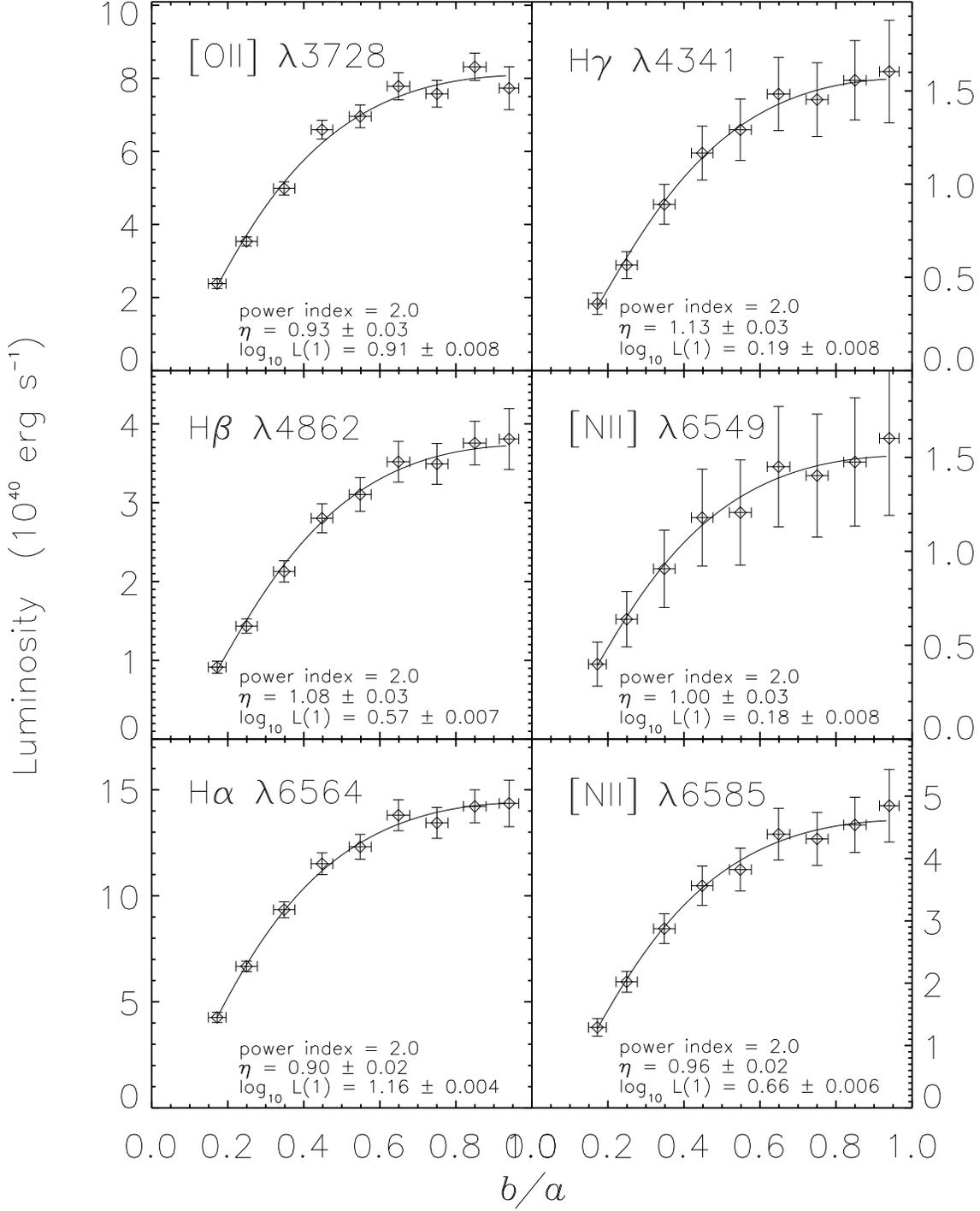}
\caption{The nebular luminosities ($\pm$ one sigma uncertainty) of the
disk galaxies  as a  function of inclination.   The solid line  is the
best-fit        relation        $\log_{10}\left[L(b/a)\right]        -
\log_{10}\left[L(1)\right] = - \eta \, \log_{10}^{2}(\ba)$. The legend
shows the best-fit coefficients.}
\label{fig:lum_vs_inclination_alllines}
\end{figure}

The  dependence  of  \halpha,  \hbeta,  \hgamma, \oii  \  and  \nii  \
luminosities       on       inclination       are       shown       in
\figname\ref{fig:lum_vs_inclination_alllines}    and    tabulated   in
\tabname\ref{tab:line}.  Their luminosities decrease with inclination.
We  deduce that  this  line-luminosity--inclination trend  is caused  by the
presence  of diffuse  dust  in  the disk.   The  reasons are  twofold.
Firstly, the  geometry of the dust  localized to the  \hiiregions \ is
not expected  to correlate with  the disk inclination.   Secondly, the
relative  extinction  in   the  line  luminosities  is  \relextHalpha,
calculated  from   the  factor  of   \luminositydropwithinclination  \
luminosity  drop in  the observed  inclination range.   This  value is
consistent       with        the       extinction       in       disks
\citep{2007MNRAS.379.1022D,2009ApJ...691..394M,2010ApJ...709..780Y},
all  being  derived  through  the inclination  dependency  of  stellar
continuum.

Fitting the luminosities with a $\log_{10}^{2}(b/a)$ dependence

\begin{equation}
\log_{10}\left[L(b/a)\right] - \log_{10}\left[L(1)\right]  = - \eta \,
\log_{10}^{2}(\ba)\label{eqn:luminosity-inclination-relation}
\end{equation}

\noindent
gives $\eta$  in the  $0.9 -1.1$ range  for these emission  lines. The
small range of $\eta$ values  is a manifestation of these lines arisen
from  same  \hiiregions.  The  ratio  of  \hiiregion  \ emissions  is,
therefore,  expected to  remain constant  with inclination  (e.g., the
constancy of Balmer decrement in \paperone).

\begin{figure}
\plotone{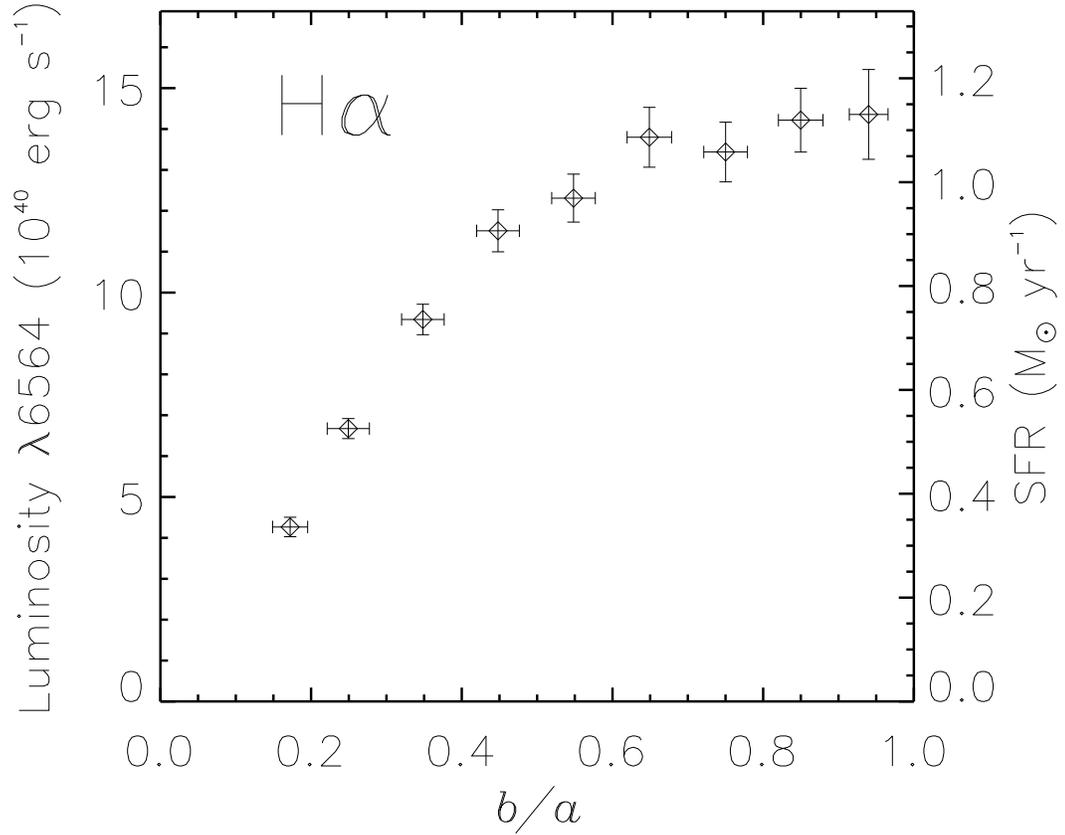}
\caption{The \halpha \ luminosity ($\pm$ one sigma uncertainty) of the
disk galaxies  as a function  of inclination.  The right  y-axis shows
the    present    \sfr,    converted   from    $L_{\halpham}$    using
\eqnname\ref{eqn:sfrha}.  The luminosity and  \sfr \ decrease by about
a factor  of \luminositydropwithinclination \ from  face-on to edge-on
orientations.}
\label{fig:lum_vs_inclination_6564.61.bayesian.gmean}
\end{figure}

\begin{figure}
\plotone{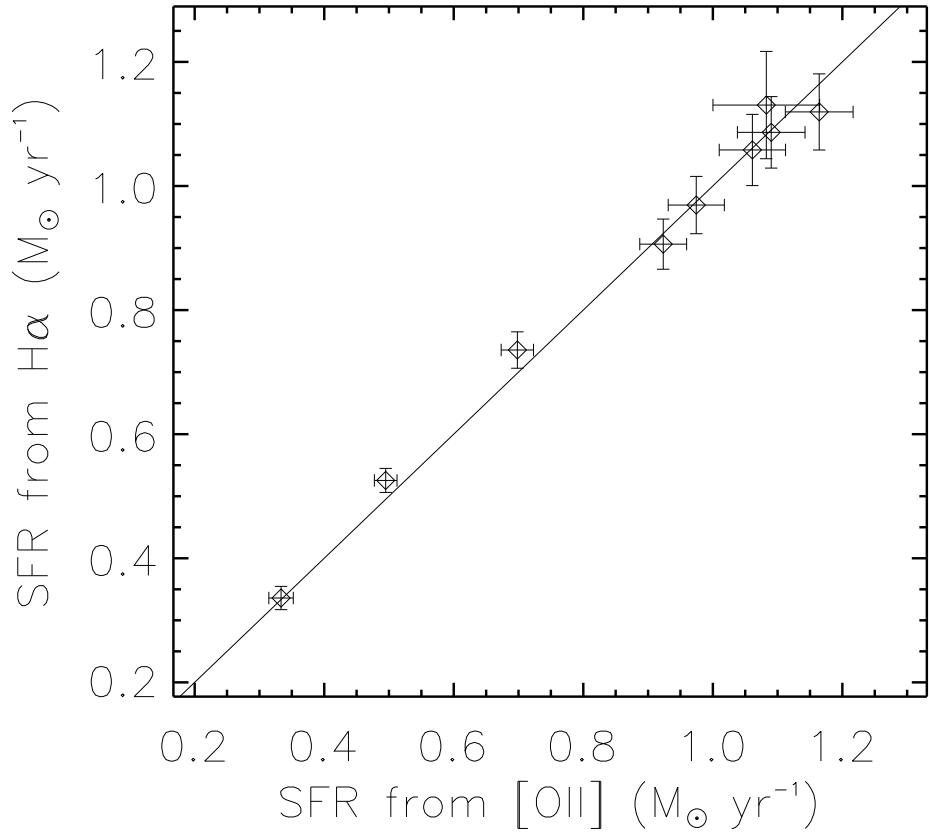}
\caption{The derived \sfr  \ using \oii \ and \halpha  \ agree well on
the inclination-to-inclination basis.}
\label{fig:line_vs_line_sfr_6564.61_vs_sfr_3728.30.bayesian.gmean}
\end{figure}

  We use the calibrations in \citet{1998ARA&A..36..189K} to derive the
present  \sfr \  due to  O/B stars  from the  line  luminosities.  The
calibrations are

\begin{equation}
\sfrmath = \frac{L_{\halpham}}{\sfrscaleha} \,\, (\solarmassperyr)
\label{eqn:sfrha}
\end{equation}

\noindent
for \halpha \ emission, and

\begin{equation}
\sfrmath = \frac{L_{\oiim}}{\sfrscaleoii} \,\, (\solarmassperyr)
\label{eqn:sfroii}
\end{equation}

\noindent 
for \oii  \ emission. The  derived \halpha \  \sfr \ are shown  on the
right                             y-axis                            of
\figname\ref{fig:lum_vs_inclination_6564.61.bayesian.gmean}.        The
average \sfr \ for our disk  galaxies is \sfraverage. The \sfr \ agree
well  in  both the  \halpha  \  and the  \oii  \  lines,  as shown  in
\figname\ref{fig:line_vs_line_sfr_6564.61_vs_sfr_3728.30.bayesian.gmean},
in the inclination-to-inclination basis. This agreement is expected if
the \halpha \  and \oii \ emissions arise from  the same \hiiregions \
and hence the same  inclination dependency.  In principle, there could
be  \ion{H}{2}  regions  embedded  within  the disks  that  cannot  be
observed.  The  nebular luminosities, and hence the  present \sfr, are
lower limits.

Combining    \eqnname\ref{eqn:luminosity-inclination-relation}    with
\eqnname\ref{eqn:sfrha}  and  \ref{eqn:sfroii},  the  present  \sfr  \
derived from \halpha \ follow this relation

\begin{equation}
\log_{10}\left[\sfrmath(1)\right] =
\log_{10}\left[\sfrmath(\ba)\right] + 0.89 \,
\log_{10}^{2}(\ba) \ ,
\end{equation}

\noindent
and similarly for the \oii-derived \sfr, 

\begin{equation}
\log_{10}\left[\sfrmath(1)\right] =
\log_{10}\left[\sfrmath(\ba)\right] + 0.91 \,
\log_{10}^{2}(\ba) \ .
\end{equation}

As in  line luminosities, the  \sfr \ decreases  by a factor  of three
from face-on  to edge-on  (axis ratio limit  being \baedgeon \  in our
sample)  galaxies.   We note  that,  although  the  derivation of  the
\sfr--inclination   relations  uses   the  correlation   between  line
luminosity and  \sfr, the  relations themselves do  not depend  on the
actual values of the proportional constants in \eqnname\ref{eqn:sfrha}
and \ref{eqn:sfroii}.

The     \oii/(\halpha    \    +     \nii)    \     luminosity    ratio
(\tabname\ref{tab:line})   is  found   here  to   be   \oiitohalpha  \
(\tabname\ref{tab:line}),   in   good    agreement   with    that   by
\citet[][\oiitohalphaKennicutt]{1992ApJ...388..310K},               and
\oiitohalphaUstoHopkins   \    higher   than   the    measurement   by
\citet[][\oiitohalphaHopkins]{2003ApJ...599..971H}.

\section{Summary} \label{section:summary}

We  present  the inclination  dependency  of  the  \lineids \  nebular
luminosities  derived from  a sample  of  disk galaxies  in the  local
universe.  The extinction  is deduced to be caused  by diffuse dust in
the    disks    with    an    amplitude   of    \relextHalpha.     The
line-luminosity--inclination  relation  can  be  used to  correct  for
extinction in \sfr \ of inclined disk galaxies.  The measuring of \oii
\  luminosity  in  sky  surveys --  BOSS  \citep{2007AAS...21113229S},
WiggleZ               \citep{2007astro.ph..1876G},              PRIMUS
\citep{2010arXiv1011.4307C}, to name a  few -- would make our approach
to  become valuable  in  studying star-forming  disk  galaxies out  to
redshift of \redshiftupperlimit.

For  the continuum-generating  stars, the  model-based  face-on (i.e.,
inclination independent)  extinction is  in the optically  thin regime
(e.g.,  \paperone).   That  for  the  \hiiregion \  emissions  is  not
determined in this work.  We are investigating this question.

\section{Acknowledgments}

We    thank    Jim~Heasley,    Julianne~Dalcanton,    Andrew~Connolly,
Sandra~Faber,       David~Koo,       Joel~Primack,      Vivienne~Wild,
St\'ephane~Charlot,   Paul~Hewitt,   Robert~Kennicutt,   Aida~Wofford,
Carl~Leitherer,   Alister~Graham,   Tim~Heckman,  Brice~M\'enard   and
Rosemary~Wyse for comments, discussions, suggestions.

This research has made use  of data obtained from or software provided
by  the US  National Virtual  Observatory, which  is sponsored  by the
National Science Foundation.

Funding for  the SDSS and SDSS-II  has been provided by  the Alfred P.
Sloan Foundation, the Participating Institutions, the National Science
Foundation, the  U.S.  Department of Energy,  the National Aeronautics
and Space Administration, the  Japanese Monbukagakusho, the Max Planck
Society,  and the Higher  Education Funding  Council for  England. The
SDSS Web Site is http://www.sdss.org/.

{}

\clearpage

\begin{table}\begin{center}
\caption{Apparent size of the disk galaxies.}
\begin{tabular}{ccccc}
\hline
broadband & 
$\beta$\tablenotemark{a} & 
$\log_{10}\left(\reffm^{1}\right)$\tablenotemark{a} & 
$\reffm^{1}$ (arc-second) & 
\rfiber/$\reffm^{1}$\tablenotemark{b} \\ 
\hline
$u$ & $0.282\pm0.010$ & $0.550\pm0.004$ & $3.552\pm0.029$ & $0.422\pm0.005$\\
$g$ & $0.278\pm0.008$ & $0.508\pm0.003$ & $3.223\pm0.021$ & $0.465\pm0.004$\\
$r$ & $0.260\pm0.007$ & $0.485\pm0.003$ & $3.058\pm0.018$ & $0.491\pm0.004$\\
$i$ & $0.240\pm0.007$ & $0.476\pm0.003$ & $2.991\pm0.018$ & $0.501\pm0.004$\\
$z$ & $0.231\pm0.007$ & $0.461\pm0.002$ & $2.889\pm0.017$ & $0.519\pm0.004$\\
\hline
\end{tabular}
\label{tab:size}
\tablecomments{Galaxy apparent size increases toward shorter wavelengths.}
\tablenotetext{a}{The  best-fit coefficients  in the  adopted relation
  between the  apparent half-light  radius and the  inclination. See \S\ref{section:fiber-to-circle}.}
  \tablenotetext{b}{The  radius  of the  SDSS spectroscopic  fibers,  \rfiber,  is  \rfibersdss.}  
\end{center}\end{table}

\begin{landscape}
\begin{table}
\caption{Emission luminosity and ratio as a function of inclination.}
\begin{tabular}{cc|rrrrrrr}
\hline
$b/a$ & number 
& [OII]$\lambda$        3728
& H$\gamma$$\lambda$        4341
& H$\beta$$\lambda$        4862
& [NII]$\lambda$        6549
& H$\alpha$$\lambda$        6564
& [NII]$\lambda$        6585
&[OII]/(H$\alpha$ + [NII])
\\
\hline
$0.94\pm0.03$ & $         429$
 & $  7.73\pm  0.59$
 & $  1.60\pm  0.28$
 & $  3.81\pm  0.39$
 & $  1.60\pm  0.41$
 & $ 14.36\pm  1.10$
 & $  4.85\pm  0.58$
 & $ 0.37\pm 0.04$
\\
$0.85\pm0.03$ & $         759$
 & $  8.31\pm  0.37$
 & $  1.56\pm  0.21$
 & $  3.76\pm  0.28$
 & $  1.48\pm  0.34$
 & $ 14.22\pm  0.78$
 & $  4.54\pm  0.44$
 & $ 0.41\pm 0.03$
\\
$0.75\pm0.03$ & $         736$
 & $  7.58\pm  0.37$
 & $  1.45\pm  0.20$
 & $  3.49\pm  0.26$
 & $  1.40\pm  0.33$
 & $ 13.44\pm  0.73$
 & $  4.31\pm  0.43$
 & $ 0.40\pm 0.03$
\\
$0.65\pm0.03$ & $         729$
 & $  7.78\pm  0.37$
 & $  1.48\pm  0.20$
 & $  3.52\pm  0.26$
 & $  1.45\pm  0.32$
 & $ 13.80\pm  0.73$
 & $  4.39\pm  0.41$
 & $ 0.40\pm 0.03$
\\
$0.55\pm0.03$ & $         783$
 & $  6.96\pm  0.31$
 & $  1.29\pm  0.16$
 & $  3.10\pm  0.21$
 & $  1.21\pm  0.28$
 & $ 12.31\pm  0.59$
 & $  3.82\pm  0.34$
 & $ 0.40\pm 0.02$
\\
$0.45\pm0.03$ & $         806$
 & $  6.59\pm  0.26$
 & $  1.17\pm  0.14$
 & $  2.80\pm  0.18$
 & $  1.18\pm  0.26$
 & $ 11.51\pm  0.51$
 & $  3.56\pm  0.32$
 & $ 0.41\pm 0.02$
\\
$0.35\pm0.03$ & $         942$
 & $  4.99\pm  0.18$
 & $  0.89\pm  0.11$
 & $  2.13\pm  0.14$
 & $  0.91\pm  0.21$
 & $  9.34\pm  0.37$
 & $  2.87\pm  0.24$
 & $ 0.38\pm 0.02$
\\
$0.25\pm0.03$ & $         828$
 & $  3.53\pm  0.12$
 & $  0.57\pm  0.07$
 & $  1.43\pm  0.09$
 & $  0.64\pm  0.15$
 & $  6.67\pm  0.25$
 & $  2.02\pm  0.17$
 & $ 0.38\pm 0.02$
\\
$0.17\pm0.02$ & $         271$
 & $  2.38\pm  0.13$
 & $  0.36\pm  0.06$
 & $  0.91\pm  0.08$
 & $  0.40\pm  0.12$
 & $  4.27\pm  0.24$
 & $  1.29\pm  0.14$
 & $ 0.40\pm 0.03$
\\
\hline
\end{tabular}
\label{tab:line}
\tablecomments{The luminosities  are in units  of $10^{40}$~\ergs,
  $\pm $ one sigma uncertainty.}
\end{table}
\end{landscape}

\clearpage

\begin{thebibliography}{}
\bibitem[Adelman-McCarthy et al.(2008)]{2008ApJS..175..297A} 
Adelman-McCarthy, J.~K., Ag{\"u}eros, M.~A., Allam, S.~S., et al.\ 2008, 
\apjs, 175, 297 
\bibitem[Athanassoula et al.(1993)]{1993A&AS..102..229A} Athanassoula, E., Garcia-Gomez, C., \& Bosma, A.\ 1993, \aaps, 102, 229 
\bibitem[Bailin \& Harris(2008)]{2008ApJ...681..225B} Bailin, J., \& Harris, W.~E.\ 2008, \apj, 681, 225 
\bibitem[Brinchmann et al.(2004)]{2004MNRAS.351.1151B} Brinchmann, J., 
Charlot, S., White, S.~D.~M., et al.\ 2004, \mnras, 351, 1151 
\bibitem[Calzetti et al.(2000)]{2000ApJ...533..682C} Calzetti, D., Armus, 
L., Bohlin, R.~C., et al.\ 2000, \apj, 533, 682 
\bibitem[Coil et al.(2010)]{2010arXiv1011.4307C} Coil, A.~L., Blanton, 
M.~R., Burles, S.~M., et al.\ 2010, arXiv:1011.4307 
\bibitem[Connolly et al.(1997)]{1997ApJ...486L..11C} Connolly, A.~J., 
Szalay, A.~S., Dickinson, M., Subbarao, M.~U., 
\& Brunner, R.~J.\ 1997, \apjl, 486, L11 
\bibitem[Conroy et al.(2010)]{2010ApJ...718..184C} Conroy, C., 
Schiminovich, D., \& Blanton, M.~R.\ 2010, \apj, 718, 184 
\bibitem[Dobos et al.(2004)]{2004ASPC..314..185D} Dobos, L., Budav{\'a}ri, 
T., Csabai, I., 
\& Szalay, A.~S.\ 2004, Astronomical Data Analysis Software and Systems (ADASS) XIII, 314, 185 
\bibitem[Driver et al.(2007)]{2007MNRAS.379.1022D} Driver, S.~P., Popescu, 
C.~C., Tuffs, R.~J., Liske, J., Graham, A.~W., Allen, P.~D., 
\& de Propris, R.\ 2007, \mnras, 379, 1022 
\bibitem[Driver et al.(2008)]{2008ApJ...678L.101D} Driver, S.~P., Popescu, 
C.~C., Tuffs, R.~J., et al.\ 2008, \apjl, 678, L101 
\bibitem[Glazebrook et al.(2007)]{2007astro.ph..1876G} Glazebrook, K., 
Blake, C., Couch, W., et al.\ 2007, arXiv:astro-ph/0701876 
\bibitem[Hakobyan et al.(2007)]{2007Ap.....50..426H} Hakobyan, A.~A., 
Petrosian, A.~R., Yeghazaryan, A.~A., \& Boulesteix, J.\ 2007, Astrophysics, 50, 426 
\bibitem[Hodge(1969)]{1969ApJ...155..417H} Hodge, P.~W.\ 1969, \apj, 155, 417 
\bibitem[Hodge \& Kennicutt(1983)]{1983ApJ...267..563H} Hodge, P.~W., \& Kennicutt, R.~C., Jr.\ 1983, \apj, 267, 563 
\bibitem[Holwerda et al.(2007)]{2007AJ....134.2385H} Holwerda, B.~W., Keel, 
W.~C., \& Bolton, A.\ 2007, \aj, 134, 2385 
\bibitem[Hopkins et al.(2003)]{2003ApJ...599..971H} Hopkins, A.~M., Miller, 
C.~J., Nichol, R.~C., et al.\ 2003, \apj, 599, 971 
\bibitem[Hopkins 
\& Beacom(2006)]{2006ApJ...651..142H} Hopkins, A.~M., \& Beacom, J.~F.\ 2006, \apj, 651, 142 
\bibitem[Huizinga 
\& van Albada(1992)]{1992MNRAS.254..677H} Huizinga, J.~E., \& van Albada, T.~S.\ 1992, \mnras, 254, 677 
\bibitem[Kennicutt(1992)]{1992ApJ...388..310K} Kennicutt, R.~C., Jr.\ 1992, 
\apj, 388, 310 
\bibitem[Kennicutt(1998)]{1998ARA&A..36..189K} Kennicutt, R.~C., Jr.\ 1998, \araa, 36, 189 
\bibitem[Koopmann et al.(2006)]{2006AJ....131..716K} Koopmann, R.~A., 
Haynes, M.~P., \& Catinella, B.\ 2006, \aj, 131, 716
\bibitem[Lilly et al.(1996)]{1996ApJ...460L...1L} Lilly, S.~J., Le Fevre, 
O., Hammer, F., \& Crampton, D.\ 1996, \apjl, 460, L1 
\bibitem[Madau et al.(1998)]{1998ApJ...498..106M} Madau, P., Pozzetti, L., 
\& Dickinson, M.\ 1998, \apj, 498, 106 
\bibitem[Maller et al.(2009)]{2009ApJ...691..394M} Maller, A.~H., Berlind, 
A.~A., Blanton, M.~R., \& Hogg, D.~W.\ 2009, \apj, 691, 394 
\bibitem[Masters et al.(2010)]{2010MNRAS.404..792M} Masters, K.~L., Nichol, 
R., Bamford, S., et al.\ 2010, \mnras, 404, 792 
\bibitem[M{\"o}llenhoff et 
al.(2006)]{2006A&A...456..941M} M{\"o}llenhoff, C., Popescu, C.~C., \& Tuffs, R.~J.\ 2006, \aap, 456, 941 
\bibitem[Ryden(2004)]{2004ApJ...601..214R} Ryden, B.~S.\ 2004, \apj, 601, 
214 
\bibitem[Schlegel et al.(2007)]{2007AAS...21113229S} Schlegel, D.~J., 
Blanton, M., Eisenstein, D., et al.\ 2007, Bulletin of the American 
Astronomical Society, 38, \#132.29 
\bibitem[Shao et al.(2007)]{2007ApJ...659.1159S} Shao, Z., Xiao, Q., Shen, 
S., Mo, H.~J., Xia, X., \& Deng, Z.\ 2007, \apj, 659, 1159 
\bibitem[Unterborn \& Ryden(2008)]{2008ApJ...687..976U} Unterborn,
  C.~T., \ \& Ryden, B.~S.\ 2008, \apj, 687, 976 
\bibitem[Wild et al.(2011)]{2011MNRAS.tmp.1545W} Wild, V., Charlot, S., 
Brinchmann, J., et al.\ 2011, \mnras, 1545 
\bibitem[Yip et al.(2010)]{2010ApJ...709..780Y} Yip, C.-W., Szalay, A.~S., 
Wyse, R.~F.~G., Dobos, L., Budav{\'a}ri, T., \& Csabai, I.\ 2010, \apj, 709, 780  (\paperone)
\bibitem[York et al.(2000)]{2000AJ....120.1579Y} York, D.~G., et al.\ 2000, 
\aj, 120, 1579
\bibitem[Zanstra(1927)]{1927ApJ....65...50Z} Zanstra, H.\ 1927, \apj, 65, 50 
\end{thebibliography}
\end{document}